\documentclass[10pt]{article}
\usepackage{verbatim}
\usepackage{amsmath,amssymb}
\makeatletter \@addtoreset{equation}{section} \makeatother

\newcommand{\noi}{\vspace{12pt}\noindent}
\newcommand{\beq}{\begin{equation}}
\newcommand{\eeq}{\end{equation}}
\newcommand{\bea}{\begin{eqnarray}}
\newcommand{\eea}{\end{eqnarray}}

\newcommand{\e}[1]{{(\ref{#1})}}
\newcommand{\eq}[1]{{eq.\ (\ref{#1})}}
\newcommand{\es}[2]{{(\ref{#1}) and (\ref{#2})}}
\newcommand{\eqs}[2]{{eqs.\ (\ref{#1}) and (\ref{#2})}}
\newcommand{\Ref}[1]{{Ref.~\cite{#1}}}

\newcommand{\equi}[1]{\stackrel{{#1}}{=}}

\newcommand{\ie}{{${ i.e., \ }$}}
\newcommand{\eg}{{${ e.g., \ }$}}

\newcommand{\cf}{{cf.\ }}

\newcommand{\wrt}{{with respect to }}

\newcommand{\wthot}{{with the help of the }}

\newcommand{\rhs}{{right--hand side }}

\renewcommand{\~}{ \ }
\renewcommand{\=}{ \ = \ }
\newcommand{\eps}{\varepsilon^{}}
\renewcommand{\tilde}{\widetilde}
\renewcommand{\bar}{\overline}

\newcommand{\for}{{\rm for}}
\newcommand{\gauged}{{\rm gauged}}

\newcommand{\Hf}{\frac{1}{2}}

\newcommand{\cL}{{\cal L}}
\newcommand{\tcL}{\widetilde{\cal L}}
\newcommand{\R}{\mathbb{R}}
\newcommand{\Rd}{\mathbb{R}^{d}}
\newcommand{\dx}{\! {\rm d}^{d}x}
\newcommand{\tx}{\widetilde{x}}
\newcommand{\tphi}{\widetilde{\phi}}
\newcommand{\cU}{{\cal U}}
\newcommand{\cV}{{\cal V}}
\newcommand{\cW}{{\cal W}}
\newcommand{\hY}{\hat{Y}}

\newcommand{\proofbox}{\begin{flushright}{\hfill \ensuremath{\Box}}
\end{flushright}}

\newtheorem{theorem}{Theorem}[section]

\newtheorem{lemma}[theorem]{Lemma}

\begin{document}
\thispagestyle{empty}
\title{\Large{\bf Noether's Theorem for a Fixed Region}}
\author{{\sc Klaus~Bering} \\~\\
Institute for Theoretical Physics \& Astrophysics\\
Masaryk University\\Kotl\'a\v{r}sk\'a 2\\CZ--611 37 Brno\\Czech Republic}
\maketitle
\vfill
\begin{abstract}
We give an elementary proof of Noether's first Theorem while stressing the
magical fact that the global quasi-symmetry only needs to hold for one fixed 
integration region. We provide sufficient conditions for gauging a global 
quasi-symmetry.  
\end{abstract}
\vfill
\begin{quote}
MSC number(s): 70S10. 
\hrule width 5.cm \vskip 2.mm \noindent 
\small E--mail:~{\tt bering@physics.muni.cz} \\
\end{quote}

\newpage

\section{Introduction}
\label{secintro}

\noi
We shall assume that the reader is familiar with Noether's Theorem in its
most basic formulation. {}For a general introduction to the subject and for 
references, see \eg Goldstein's book \cite{goldstein80} and the Wikipedia
entry for Noether's Theorem \cite{wiki}. The purpose of this paper is to state
and prove Noether's Theorem in a powerful field-theoretic setting with a
minimum of assumptions. At the same time, we aim at being self-contained and
using as little mathematical machinery as practically possible.

\noi
Put into one sentence, the first Theorem of Noether states that a continuous,
global, off-shell quasi-symmetry of an action $S$ implies a local on-shell 
conservation law, \ie a continuity equation for a Noether current, which is 
valid in each world-volume point. Strictly speaking, Noether herself
\cite{noether18} and the majority of authors talk about {\em 
symmetry/invariance} rather than {\em quasi-symmetry/quasi-invariance}, but 
since quasi-symmetry is a very useful, natural and relatively mild
generalization, we shall only use the notion of quasi-symmetry here, \cf 
Section~\ref{secsym}. The term {\em global} is defined in 
Section~\ref{secglobal}.

\noi
The traditional treatment of Noether's first Theorem assumes that the global 
quasi-symmetry of the action $S$ holds for {\em every} integration region, see
\eg Noether \cite{noether18}, Hill \cite{hill51}, Goldstein \cite{goldstein80},
Bogoliubov and Shirkov \cite{bogoliubovshirkov80}, Trautman \cite{trautman67},
Komorowski \cite{komorowski68}, Ibragimov \cite{ibragimov69}, Sarlet and 
Cantrijn \cite{sarletcantrijn81}, Olver \cite{olver86}, and Ramond
\cite{ramond89}. In the case of Olver \cite{olver86}, this assumption is hidden
inside his definition of symmetry. Adding to the confusion, Goldstein
\cite{goldstein80} and Ramond \cite{ramond89} do never explicitly state that
they require the quasi-symmetry of the action $S$ to hold for {\em every}
integration region, but this is the only interpretation that is consistent with
their further conclusions, technically speaking, because their Noether identity
contains only the bare (rather than the improved) Noether current. 

\noi
There is also a non-integral version of Noether's Theorem based on a 
quasi-symmetry of the Lagrangian density $\cL(x)$ (or the Lagrangian form 
$\cL(x){\rm d}^{d}x$) rather than the action $S$, see \eg Arnold 
\cite{arnold89}, or Jos\'e and Saletan \cite{josesaletan98}. We shall here 
only discuss integral formulations.

\begin{table}[ht]
\caption{Flow-diagram of Noether's first Theorem.
The $J^{\mu}(x)$ in Table~\ref{flowtable1} is an (improved) Noether current, 
\cf Section~\ref{secsym}, and $Y^{\alpha}_{0}$ is a vertical generator of 
quasi-symmetry, see Section~\ref{secvar}. The term {\em on-shell} and the wavy
equality sign ``$\approx$'' means that the equations of motion
$\delta\cL(x)/\delta\phi^{\alpha}(x)\approx 0$ has been used. 
}

\label{flowtable1}
\begin{center}
\begin{tabular}{c}
Continuous global off-shell quasi-symmetry of \\
$S^{}_{\cV}=\int_{\cV}\dx\~\cL(x)$ for a fixed region $\cV$. \\ 
$\Downarrow$ \\ 
Continuous global off-shell quasi-symmetry of \\
$S^{}_{\cU}=\int_{\cU}\dx\~\cL(x)$ for every region $\cU\subseteq \cV$. \\ 
$\Downarrow$  \\ 
Local off-shell Noether identity: \\
$\forall \phi:\~\~ d_{\mu}J^{\mu}(x)
\equiv -\frac{\delta \cL(x)}{\delta \phi^{\alpha}(x)} Y^{\alpha}_{0}(x)$\~.\\ 
$\Downarrow$ \\ 
Local on-shell conservation law:\\ 
$d_{\mu}J^{\mu}(x) \approx 0$\~. 
\end{tabular}
\end{center}
\end{table}

\noi
If the action $S$ has quasi-symmetry for {\em every} integration region, it is,
in retrospect, not surprising that one can derive a {\em local} conservation
law for a Noether current via localization techniques, \ie by chopping the
integral $S$ into smaller and smaller neighborhoods around a single
world-volume point. It would be much more amazing, if one could derive a {\em
local} conservation law from only the knowledge that the action $S$ has a 
quasi-symmetry for {\em one fixed} integration region. Our main goal with this
paper is to communicate to a wider 
audience that this is possible! More precisely, the statement is, firstly, that
the global quasi-symmetry of the action $S$ only needs to hold for {\em one
fixed} region of the world volume, namely the pertinent full world volume $\cV$,
and secondly, that this will, in turn, imply a global quasi-symmetry for {\em 
every} smaller region $\cU\subseteq \cV$. (We assume that the target space $M$
is contractible, \cf Section~\ref{secwvts}, and that the quasi-symmetry
is projectable, \cf Section~\ref{secvar}.) It is for aesthetic and practical
reasons nice to minimize the assumptions, and when formulated with a fixed
region, the conclusions in Noether's first Theorem are mesmerizingly strong,
\cf Table~\ref{flowtable1}. The crucial input is the strong assumption that the
quasi-symmetry of $S$ should be valid {\em off-shell}, \ie for {\em every} 
possible configurations of the field $\phi$; not just for configurations that 
satisfy equations of motion.
To our knowledge, a proof of these facts has not been properly written down
anywhere in the literature in elementary terms, although the key idea is
outlined by, \eg Polchinski \cite{polchinski98}. (See also de Wit and Smith
\cite{dewitsmith86}.)

\noi
The paper is organized as follows. The main proof and definitions are given in 
Sections~\ref{secwvts}--\ref{secsym}, while Section~\ref{secgenmfld} and 
Appendix~\ref{secapp} provide some technical details. 
Sections~\ref{secex}--\ref{secex3} contain examples from classical mechanics of
a global, off-shell, symmetry \wrt one fixed region that is {\em not} a
symmetry for generic regions. {}Finally, Appendix~\ref{secgauge} provides 
closed formulas and sufficient conditions for gauging a global quasi-symmetry.

\section{World Volume and Target Space}
\label{secwvts}

\noi
Consider a field $\phi:\cV\!\to\! M$ from a fixed $d$-dimensional world volume 
$\cV$ to a target space $M$. (We use the term {\em world volume} rather than the
more conventional term {\em space-time}, because space-time in, \eg string
theory is associated with the target space.) We will first consider the special
case where $\cV\subseteq\Rd$, and postpone the general case where $\cV$ is a 
general manifold to Section~\ref{secgenmfld}. Here $\R$ denotes the set of real
numbers. We will always assume for simplicity that the target space $M$ has 
global coordinates $y^{\alpha}$, so that one can describe the field $\phi$ with
its coordinate functions $y^{\alpha}\!=\!\phi^{\alpha}(x)$, $x\in \cV$. We
furthermore assume that the $y^{\alpha}$-coordinate region (which we identify
with the target space $M$) is {\em star-shaped} around a point (which we take
to be the origin $y\!=\!0$),
\ie 
\beq
\forall y \in M \forall \lambda \in [0,1]:\~\~\lambda y \in M\~.
\label{starshaped}
\eeq
The world volume $\cV$ and the target space $M$ are also called the {\em 
horizontal} and the {\em vertical space}, respectively.

\section{Action $S^{}_{\cV}$}

\noi
The {\em action} $S^{}_{\cV}$ is given as a local functional
\beq
S^{}_{\cV}[\phi]\~:=\~\int_{\cV}\dx\~\cL(x)  \label{action0}
\eeq
over the world volume $\cV$, where the {\em Lagrangian density}
\beq
\cL(x)\=\cL(\phi(x),\partial\phi(x),x) \label{lagrdensity0}
\eeq
depends smoothly on the fields $\phi^{\alpha}(x)$, their first derivatives 
$\partial^{}_{\mu}\phi^{\alpha}(x)$, and explicitly on the point $x$. Phrased
mathematically, the Lagrangian density
$\cL \in C^{\infty}(M\!\times\! M^{d}\!\times\! \cV)$ is assumed to be a smooth
function on the $1$-jet space. Please note that the $\phi$ and the
$\partial\phi$ dependence will often not be written explicitly, cf., \eg the 
\rhs of \eq{action0}.
Since we do not want to impose boundary conditions on the field $\phi(x)$ (at 
least not at this stage), the notion of functional/variational derivative 
$\delta S^{}_{\cV} / \delta\phi(x)$ may be ill-defined, see \eg \Ref{b00}. 
In contrast, the Euler-Lagrange derivative $\delta\cL(x)/\delta\phi(x)$ is
always well-defined, \cf \eq{eom}, even if the principle of stationary/least 
action has an incomplete formulation (at this stage). So when we speak of {\em 
equations of motion} and {\em on-shell}, we mean the equations 
$\delta\cL(x)/\delta\phi(x)\approx0$. (We should finally mention that
Noether's Theorem also holds if the Lagrangian density $\cL$ contains higher 
derivatives $\partial^{2}\phi$, $\partial^{3}\phi$, $\ldots$, 
$\partial^{n}\phi$, of the field $\phi$, and/or if the world volume $\cV$
and/or if the target space $M$ are supermanifolds, but we shall for simplicity
not consider this here.)

\noi
We will consider three cases of the fixed world volume $\cV$.
\begin{enumerate}
\item
\underline{Case $\cV=\Rd$}: The reader who does not care about subtleties 
concerning boundary terms can assume $\cV\!=\!\Rd$ from now on (and ignore hats 
``$\wedge$'' on some symbols below).

\item
\underline{Case $\cV \subset \Rd$}: {}For notational reasons it is convenient to
assume that the original Lagrangian density
$\cL\in C^{\infty}(M\!\times\! M^{d} \!\times\! \cV)$ in \eq{action0} and every 
admissible field configuration $\phi:\cV\!\to\!M$ can be smoothly extended 
to some function $\cL\in C^{\infty}(M\!\times\! M^{d} \!\times\! \Rd)$ and to 
functions $\phi:\Rd\!\to\!M$, which, with a slight abuse of notation, are 
called by the same names, respectively. The construction will actually not 
depend on which such smooth extensions are used, as will become evident 
shortly. Then it is possible to write the action \e{action0} as an integral 
over the whole $\Rd$. 
\beq
S^{}_{\cV}[\phi]\=\int_{\Rd}\dx\~\hat{\cL}(x)\~, \qquad
\hat{\cL}(x)\~:=\~1^{}_{\cV}(x)\cL(x)\~,\label{actionrd}
\eeq
where
\beq
1^{}_{\cV}(x)\~:=\~\left\{\begin{array}{ccc} 1 &\for& x \in \cV\~, \cr 
0 &\for& x \in \Rd\backslash \cV\~, \end{array}\right.  
\eeq
is the {\em characteristic function} for the region $\cV$ in $\Rd$. Note that 
$1^{}_{\cV}:\Rd\!\to\!\R$ and $\hat{\cL}:M\!\times\! M^{d}\!\times\!\Rd\!\to\!\R$
are {\em not} continuous functions. It is necessary to impose a
regularity condition for the boundary $\partial \cV$ of the region $\cV$. 
Technically, the boundary $\partial \cV\!\subset\!\Rd$ should have Lebesgue 
measure zero.
 
\item
\underline{Case $\cV$ is a general manifold}: See Section~\ref{secgenmfld}.
\end{enumerate}

\begin{table}[ht]
\caption{Conversion between notation used by various authors.}

\label{tablgoldstein}
\begin{center}
\begin{tabular}{|l||c|c|c|c|c|c|}  \hline
&Noe-&Hill&Gold-&Bogoliu-&Ra-&This \\
&ther&\cite{hill51}&stein&bov \& &mond&paper \\
&\cite{noether18}&&\cite{goldstein80}
&Shirkov \cite{bogoliubovshirkov80}&\cite{ramond89}& \\ \hline\hline
Action&$I$&$J$&$I$&${\cal A}$&$S$&$S$\rule[-1.5ex]{0ex}{4.5ex}  \\ \hline 
Lagrangian density&$f$&$\cL$&$\cL$&$\cL$&$\cL$&$\cL$\rule[-1.5ex]{0ex}{4.5ex} 
\\ \hline 
{}Field&$u^{}_{i}$&$\psi^{\alpha}$&$\eta^{}_{\rho}$&$u^{}_{i}$&$\Phi$
&$\phi^{\alpha}$\rule[-1.5ex]{0ex}{4.5ex} \\ \hline
Region&&$R$&$\Omega$&&&$\cV$\rule[-1.5ex]{0ex}{4.5ex}  \\ \hline
Infinitesimal variation&$\Delta$, $\delta$&$\delta$&$\delta$&$\delta$&$\delta$
&$\delta$\rule[-1.5ex]{0ex}{4.5ex}\\ \hline 
Vertical variation&$\bar{\delta}$&$\delta^{}_{\ast}$&$\bar{\delta}$
&$\bar{\delta}$&$\delta^{}_{0}$&$\delta^{}_{0}$
\rule[-1.5ex]{0ex}{4.5ex}\\ \hline  
Generator&&$\eta^{\alpha}$&$\Psi^{}_{\rho}$&$\Psi^{}_{i}$&&$Y^{\alpha}$
\rule[-1.5ex]{0ex}{4.5ex} \\ \hline  
Euler-Lagrange deriv.&$\psi^{}_{i}$&$[\cL]^{}_{\alpha}$&&&
&$\frac{\delta\cL(x)}{\delta\phi^{\alpha}(x)}$ 
\rule[-1.5ex]{0ex}{4.5ex}  \\ \hline
Bare Noether current &$-B$&&&$-\theta^{i}$&$-\jmath^{\mu}$&$\jmath^{\mu}$
\rule[-1.5ex]{0ex}{4.5ex} \\ \hline  
\end{tabular}
\end{center}
\end{table}

\section{Total derivative $d_{\mu}$} 

\noi
The {\em total derivative} $d_{\mu}$ is an explicit derivative $\partial_{\mu}$
plus implicit differentiation through $\phi$, $\partial \phi^{\alpha}$, 
$\ldots$, \ie
\beq
d_{\mu}\=\partial_{\mu}
+\phi^{\alpha}_{\mu}(x)\frac{\partial}{\partial \phi^{\alpha}(x)}
+\phi^{\alpha}_{\mu\nu}(x)\frac{\partial}{\partial \phi^{\alpha}_{\nu}(x)}
+\ldots\~,
\eeq
where the following shorthand notation is used
\bea
d_{\mu}&:=&\frac{d}{dx^{\mu}}\~,\qquad 
\partial_{\mu}\~:=\~\frac{\partial}{\partial x^{\mu}}\~,\qquad \cr
\phi^{\alpha}_{\mu}(x)&:=&\partial_{\mu}\phi^{\alpha}(x) \~,\qquad 
\phi^{\alpha}_{\mu\nu}(x)
\~:=\~\partial_{\mu}\partial_{\nu}\phi^{\alpha}(x)\~,\qquad \ldots\~.
\eea

\section{Variation of $x$, $\phi$ and $\cV$}
\label{secvar}

\noi
We will assume that the reader is familiar with the notion of infinitesimal
variations in a field-theoretic context. See \eg Goldstein 
\cite{goldstein80}, \cf Table~\ref{tablgoldstein}. Consider an infinitesimal 
variation $\delta$ of the coordinates $x^{\mu}\to x^{\prime\mu}$, of the fields
$\phi^{\alpha}(x)\to\phi^{\prime \alpha}(x')$, and of the region
$\cV\to \cV^{\prime}:=\{x^{\prime}\mid x \in \cV\}$, \ie
\beq
\begin{array}{c}
\begin{array}{lclcl}
x^{\prime \mu}-x^{\mu}&=:& \delta x^{\mu}&=&X^{\mu}(x)\eps(x)
\~, \cr
\phi^{\prime \alpha}(x^{\prime})-\phi^{\alpha}(x)
&=:&\delta\phi^{\alpha}(x)&=&Y^{\alpha}(x)\eps(x)\~,\cr
\phi^{\prime \alpha}(x)-\phi^{\alpha}(x)
&=:&\delta^{}_{0}\phi^{\alpha}(x)&=&Y_{0}^{\alpha}(x)\eps(x)\~,\cr
d^{\prime}_{\mu}\phi^{\prime \alpha}(x^{\prime})-d_{\mu}\phi^{\alpha}(x)
&=:&\delta d_{\mu}\phi^{\alpha}(x)&\neq&d_{\mu}\delta\phi^{\alpha}(x)\~,\cr
d_{\mu}\phi^{\prime \alpha}(x)-d_{\mu}\phi^{\alpha}(x)
&=:&\delta^{}_{0}d_{\mu}\phi^{\alpha}(x)
&=&d_{\mu}\delta^{}_{0}\phi^{\alpha}(x)\~,
\end{array}\cr
\begin{array}{c}
X^{\mu}(x) {\rm~and~} \eps(x) 
{\rm~are~independent~of~}\phi~
(\rm also~known~as~{\it projectable} \cite{olver86})
\~, \\
Y^{\alpha}(x)\=Y^{\alpha}(\phi(x),\partial\phi(x),x)\~,\qquad
Y_{0}^{\alpha}(x)\=Y_{0}^{\alpha}(\phi(x),\partial\phi(x),x)\~,
\end{array}
\end{array}
\label{variation} 
\eeq
where $\eps:\cV\!\to\! \R$ is an arbitrary infinitesimal function, and where
$X^{\mu},Y^{\alpha},Y_{0}^{\alpha}\in C^{\infty}(M\!\times\!M^{d}\!\times\!\cV)$
are {\em generators} of the variation, and in differential-geometric terms, 
they are {\em vector fields}. 

\noi
(While working with infinitesimal quantities has intuitive advantages, it
requires a comment to make them mathematically well-defined. The 
$\eps$-function should more correctly be viewed as a product
$\eps(x)\!=\!\eps_{0}h(x)$, where $\eps_{0}$ is the underlying
$1$-parameter of the variation, and $h(x)$ is a function. A {\em $1$-parameter}
means a $1$-dimensional parameter. Then, for instance, the
first equation in \e{variation} should more properly be written
$x^{\prime \mu}\!-\!x^{\mu}\!=\!\eps(x)X^{\mu}(x)\!+\!\eps_{0}o(1)$, where the
little-o notation $o(1)$ means any function of $\eps_{0}$ that vanishes in the 
limit $\eps_{0}\!\to\!0$. We shall not write such $o(1)$ terms explicitly to 
avoid clutter. The term $\eps_{0}o(1)$ is also written as $o(\eps_{0})$ in the 
little-o notation. An alternative method is to view $\eps_{0}$ as an exterior
$1$-form, so that the square $\eps_{0}\!\wedge\!\eps_{0}\!=\!0$ vanishes.)

\noi
In the case $\cV\!\subset\!\Rd$, the above functions are for notational reasons
assumed to be smoothly extended to $\eps:\Rd \!\to\! \R$ and 
$X^{\mu},Y^{\alpha},Y_{0}^{\alpha}\in 
C^{\infty}(M\!\times\!M^{d}\!\times\!\Rd)$,
which, with a slight abuse of notation, are called by the same names, 
respectively. (Again the choice of extensions will not matter.)
The generator $Y^{\alpha}(x)$ can be decomposed in a vertical and a horizontal 
piece,
\beq
\delta\=\delta^{}_{0}+\delta x^{\mu} d_{\mu}\~, \qquad
Y^{\alpha}(x)\=Y_{0}^{\alpha}(x)+\phi^{\alpha}_{\mu}(x)X^{\mu}(x)\~.
\eeq
In other words, only the vertical and horizontal generators, $Y_{0}^{\alpha}$ 
and $X^{\mu}$, respectively, are independent generators of the variation 
$\delta$. The variation $\delta \cV$ of the region $\cV$ is by definition
completely specified by the horizontal part $X^{\mu}$. The main property of the
vertical variation $\delta^{}_{0}$ that we need in the following, is that it
commutes ($[\delta^{}_{0},d_{\mu}]\!=\!0$) with the total derivative $d_{\mu}$. 
This should be compared with the fact that in general 
$[\delta,d_{\mu}]\!\neq\!0$.

\noi
(In the case of Noether's second Theorem and local gauge symmetry, the
generators $X^{\mu}, Y^{\alpha}, Y_{0}^{\alpha}$ in \eq{variation} could in
general be differential operators that act on $\eps(x)$, but since we are here
only interested in Noether's first theorem, and ultimately letting $\eps(x)$
be an $x$-independent constant $\eps_{0}$, \cf \eq{globalvar}, such
differential operators will not contribute, so we will here for simplicity
assume that the generators $X^{\mu}, Y^{\alpha}, Y_{0}^{\alpha}$ are just
functions.)

\section{Variation of $S^{}_{\cV}$}

\noi 
The infinitesimal variation $\delta S^{}_{\cV}$ of the action $S^{}_{\cV}$ comes 
in general from four types of contributions:
\begin{itemize}
\item
Variation of the Lagrangian density $\cL(x)$.
\beq
\delta\cL(x)
\=\cL(\phi^{\prime}(x^{\prime}),\partial^{\prime}\phi^{\prime}(x^{\prime}),
x^{\prime})-\cL(\phi(x),\partial\phi(x),x)\~.
\eeq
\item
Variation of the measure ${\rm d}^{d}x$, which leads to a Jacobian factor.
\beq
\delta{\rm d}^{d}x\={\rm d}^{d}x^{\prime}-{\rm d}^{d}x\= 
{\rm d}^{d}x\~d_{\mu}\delta x^{\mu}\~.
\eeq
\item
Boundary terms at $|x|\!=\!\infty$. In the way we have set up the action 
\e{actionrd} on the whole $\Rd$ there are no boundary contributions 
at $|x|\!=\!\infty$ in both case 1 and 2. 
\item
Variation of the characteristic function $1^{}_{\cV}(x)$. The characteristic
function $1^{}_{\cV}(x)$ is invariant under the variation, due to a compensating
variation $\delta \cV$ of the region $\cV$.
\beq
  \delta 1^{}_{\cV}(x)\=1^{}_{\cV^{\prime}}(x^{\prime})-1^{}_{\cV}(x)\=0\~.
\eeq 
\end{itemize}

\noi 
An arbitrary infinitesimal variation $\delta S^{}_{\cV}$ of the action
$S^{}_{\cV}$ therefore consists of the two first contributions.
\bea
\delta S^{}_{\cV} 
&=&\int_{\cV^{\prime}}\!{\rm d}^{d}x^{\prime}\~\cL(\phi^{\prime}(x^{\prime}),
\partial^{\prime}\phi^{\prime}(x^{\prime}),x^{\prime})
-\int_{\cV}\dx\~\cL(\phi(x),\partial\phi(x),x) \cr
&=&\int_{\cV}\dx \left[ \delta\cL(x)
+ \cL(x)d_{\mu} \delta x^{\mu}\right]
\= \int_{\cV}\dx \left[ \delta^{}_{0}\cL(x)
+d_{\mu}( \cL(x) \delta x^{\mu})\right] \cr
&=& \int_{\cV}\dx \left[ \frac{\delta\cL(x)}{\delta\phi^{\alpha}(x)}
\delta^{}_{0}\phi^{\alpha}(x)
+d_{\mu}\left( \frac{\partial\cL(x)}{\partial\phi^{\alpha}_{\mu}(x)}
\delta^{}_{0}\phi^{\alpha}(x)+ \cL(x) \delta x^{\mu}\right)\right] \cr
&=& \int_{\cV}\dx \left[ f(x) \eps(x) 
+ \jmath^{\mu}(x) d_{\mu} \eps(x) \right]\~. \label{svar}
\eea
Here $\delta\cL(x)/\delta\phi^{\alpha}(x)$ is the Euler-Lagrange derivative
\beq
\frac{\delta\cL(x)}{\delta\phi^{\alpha}(x)}
\~:=\~\frac{\partial\cL(x)}{\partial\phi^{\alpha}(x)}
-d_{\mu}\frac{\partial\cL(x)}{\partial\phi^{\alpha}_{\mu}(x)}
\={\rm function}(\phi(x),\partial\phi(x),\partial^{2}\phi(x),x)\~,\label{eom}
\eeq
\ie the equations of motion are of at most second order. 
In equation \e{svar} we have defined the {\em bare Noether current} as
\beq
\jmath^{\mu}(x)\~:=\~
\frac{\partial\cL(x)}{\partial\phi^{\alpha}_{\mu}(x)}
Y_{0}^{\alpha}(x) + \cL(x) X^{\mu}(x)
\=\jmath^{\mu}(\phi(x),\partial\phi(x),x)\~,\label{nnoetherhat}
\eeq
and a function
\beq
f(x)\~:=\~
\frac{\delta\cL(x)}{\delta\phi^{\alpha}(x)}Y_{0}^{\alpha}(x)
+d_{\mu}\jmath^{\mu}(x)
\=f(\phi(x),\partial\phi(x),\partial^{2}\phi(x),x)\~.\label{eff}
\eeq
In differential-geometric terms,
\beq
\jmath^{\mu}(x)\~\~\rightarrow\~\~  
\bar{\jmath}^{\nu}(\bar{x})
\=\frac{\jmath^{\mu}(x)}{\det\frac{\partial \bar{x}}{\partial x}} 
\frac{\partial \bar{x}^{\nu}}{\partial x^{\mu}} 
\qquad {\rm and} \qquad
f(x)\~\~\rightarrow\~\~ \bar{f}(\bar{x})
\=\frac{f(x)}{\det\frac{\partial \bar{x}}{\partial x}} \label{jftransf} 
\eeq
behave as a density-valued vector-field and a density
under passive coordinate transformations 
$x^{\mu}\rightarrow \bar{x}^{\nu}=\bar{x}^{\nu}(x)$, respectively.

\section{Global Variation}
\label{secglobal}

\noi
The variation \e{variation} is by definition called {\em global} (or 
{\em rigid}) if
\beq
\eps(x)\=\eps_{0} \label{globalvar}
\eeq 
is an $x$-independent infinitesimal $1$-parameter. Except for
Appendix~\ref{secgauge}, let us from now on specialize the variation
\e{variation} to the global type \e{globalvar}. Then \eq{svar} becomes
\beq
\delta S^{}_{\cV} \= \eps_{0} F^{}_{\cV}\~,\qquad  
{}F^{}_{\cV}[\phi]\~:=\~\int_{\cV}\dx\~f(x)\~.
\label{globalds}
\eeq

\section{Smaller Regions $\cU\subseteq \cV$}
\label{secsmaller}

\noi
Note that $\jmath^{\mu}(x)$ and $f(x)$, from \eqs{nnoetherhat}{eff}, 
respectively, are both independent of the region $\cV$ 
in the sense that if one had built the action 
\beq
S^{}_{\cU}[\phi]\~:=\~\int_{\cU}\dx\~\cL(x)   \label{action0u}
\eeq
from a smaller region $\cU\subseteq \cV$, and smoothly extended the pertinent
functions to $\Rd$ as in \eq{actionrd}, one would have arrived at another set 
of functions $\jmath^{\mu}(x)$ and $f(x)$, that would agree with the previous 
ones within the smaller region $x\in\cU$. Similar to \eq{globalds}, the 
corresponding global variation $\delta S^{}_{\cU}$ is just
\beq
\delta S^{}_{\cU} \= \eps_{0} F^{}_{\cU}\~,\qquad  
{}F^{}_{\cU}[\phi]\=\int_{\cU}\dx \~f(x)\~,\qquad \cU\subseteq \cV \~. 
\label{bigfu}
\eeq

\section{Quasi-Symmetry}
\label{secsym}

\noi
We will in the following use again and again the fact that an integral is a 
boundary integral if and only if its Euler-Lagrange derivative vanishes, 
\cf Appendix~\ref{secapp}. 
Assume that for a fixed region $\cV$, the action $S^{}_{\cV}$ has an off-shell
quasi-symmetry under a global variation (\ref{variation}, \ref{globalvar}). 
By definition, a global off-shell {\em quasi-symmetry} means that the
corresponding infinitesimal variation $\delta S^{}_{\cV}$ of the action is an
integral over a smooth function
$g(x)=g(\phi(x),\partial\phi(x),\partial^{2}\phi(x),\ldots,x)$, \ie 
\beq
\forall \phi:\~\~\~\~\delta S^{}_{\cV}\~\equiv\~\eps_{0} \int_{\cV}\dx\~g(x)\~, 
\label{dszero}
\eeq
where   
\beq
\begin{array}{c}
g(x){\rm~is~locally~a~divergence}: \cr
\forall x^{}_{0}\in \cV \exists{\rm~local~}x^{}_{0}{\rm~neighborhood~}
\cW\subseteq \cV, \cr
\exists g^{\mu}(x)
=g^{\mu}(\phi(x),\partial\phi(x),\partial^{2}\phi(x),\ldots,x)
\forall x\in \cW:\~\~ g(x)\=d_{\mu}g^{\mu}(x)\~.
\end{array} \label{gdiv}
\eeq
The integrand $g$ is allowed to also depend on a finite number of higher
derivatives $\partial^{2}\phi$, $\partial^{3}\phi$, $\ldots$, of the field
$\phi$. As usual we assume that the function $g$ can be extended smoothly
to $\Rd$. In differential-geometric terms, the $g$ function
behaves as a density under passive coordinate transformations 
$x^{\mu}\rightarrow \bar{x}^{\nu}=\bar{x}^{\nu}(x)$.
It follows that $\int_{\cV}\dx\~g(x)$ is a boundary integral with
identically vanishing Euler-Lagrange derivative
\beq
\frac{\delta g(x)}{\delta\phi^{\alpha}(x)}\~\equiv\~ 0\~. \label{gzero}
\eeq
(One of the aspects of Noether's Theorem, that we suppress in this note for
simplicity, is the full Lie group $G$ of quasi-symmetries. We only treat
{\em one} infinitesimal quasi-symmetry at a time, \cf the $1$-parameter
$\eps_{0}$. Thus we will also only derive {\em one} conservation law at a time. 
Technically speaking, the only remnant of $G$, that is treated here, is a 
$u(1)$ Lie subalgebra.)

\noi
A quasi-symmetry is promoted to a {\em symmetry}, if
$\delta S^{}_{\cV}\equiv 0$. (It is natural to ask if it is always possible to
turn a quasi-symmetry into a symmetry by modifying the action
$\delta S^{}_{\cV}$ with a boundary integral? The answer is in general no, see
Section~\ref{secex3} for a counterexample. Thus the notion of quasi-symmetry is
an essential generalization of the original notion of symmetry discussed by
Noether \cite{noether18}.)

\noi
The variational formula \e{globalds} together with the definition \e{dszero}
of a quasi-symmetry yield
\beq
\forall \phi:\~\~\~\~
\int_{\cV}\dx\~f(x)\~\equiv\~F^{}_{\cV}[\phi]\~\equiv\~\int_{\cV}\dx\~g(x)\~.
\label{fzero}
\eeq
Now define the zero functional 
\beq
\forall \phi:\~\~\~\~
Z^{}_{\cV}[\phi]\~\equiv\~\int_{\cV}\dx\~(f-g)(x)\~\equiv\~0\~.\label{zzero}
\eeq
By performing an arbitrary variation $\delta \phi(x)$ with support in the
interior $x\in\cV^{\circ}$ of $\cV$ away from any boundaries, one concludes
that the Euler-Lagrange derivative $\delta (f-g)(x)/\delta\phi^{\alpha}(x)$
must vanish identically in the bulk $x\in \cV^{\circ}$ (=the interior of $\cV$),
\beq
\forall \phi\forall x\!\in\!\cV^{\circ}:\~\~\~\~
\frac{\delta f(x)}{\delta\phi^{\alpha}(x)}
\~\equi{\e{gzero}}\~\frac{\delta (f-g)(x)}{\delta\phi^{\alpha}(x)}
\=0\~, \label{fdfzero}
\eeq
And by continuity, $\delta f(x)/\delta\phi^{\alpha}(x)$ must vanish for all
$x\in \cV$. Lemma~\ref{lemma1} in Appendix~\ref{secapp} now yields the
following.
\beq
\begin{array}{c}
{\rm~The~integrand~} f(x){\rm~is~locally~a~divergence}: \cr
\forall x^{}_{0}\in \cV \exists{\rm~local~}x^{}_{0}{\rm~neighborhood~}
\cW\subseteq \cV, \cr
\exists f^{\mu}(x)
=f^{\mu}(\phi(x),\partial\phi(x),\partial^{2}\phi(x),x)
\forall x\in \cW:\~\~ f(x)\=d_{\mu}f^{\mu}(x)\~.
\end{array}
\label{fdivhat}
\eeq
Equations \e{bigfu}, \es{dszero}{gdiv} then imply that the global variation is 
an off-shell quasi-symmetry of the action $S^{}_{\cU}$ for all smaller regions 
$\cU\subseteq \cV$, which is one of the main conclusions. One can locally
define an {\em improved Noether current} as
\beq
J^{\mu}(x)\~:=\~\jmath^{\mu}(x)-f^{\mu}(x)
=J^{\mu}(\phi(x),\partial\phi(x),\partial^{2}\phi(x),x)\~.
\eeq
Equation \e{eff} then immediately yields the sought--for off--shell 
Noether identity \e{offshellnihat}.

\begin{theorem}[Local Off--shell Noether identity]
A continuous, global, off-shell quasi-symmetry
(\ref{variation}, \ref{globalvar}, \ref{dszero}) of an $S^{}_{\cV}$ action
\e{action0} implies a local off--shell Noether identity
\beq
d_{\mu}J^{\mu}(x)\=d_{\mu}\jmath^{\mu}(x)-f(x)
\~\equi{\e{eff}}\~
-\frac{\delta\cL(x)}{\delta\phi^{\alpha}(x)}Y_{0}^{\alpha}(x)\~.
\label{offshellnihat}
\eeq
\end{theorem}

\section{Case 3: General Manifold $\cV$}
\label{secgenmfld}

\noi
If the world volume $\cV$ is a manifold, one decomposes
$\cV=\sqcup_{a}\cV^{}_{a}$ in a disjoint union of coordinate patches. (Disjoint
modulo zero Lebesgue measure of pertinent boundaries.) Under an infinitesimal
variation \e{variation}, the  world volume transforms
$\cV\rightarrow \cV^{\prime}=\sqcup_{a}\cV^{\prime}_{a}$, where 
$\cV^{\prime}_{a}:=\{x^{\prime}\mid x \in \cV_{a}\}$. Each coordinate patch
$\cV^{}_{a}$ and its variation $\cV^{\prime}_{a}$ are identified with subsets
$\subseteq\Rd$. The $S^{}_{\cV}$ action \e{action0} decomposes
\beq
S^{}_{\cV}\=\sum_{a}S^{}_{a}\~,\qquad S^{}_{a}[\phi]\=\int_{\cV^{}_{a}}\dx\~
\cL^{}_{a}(x) \~,\qquad 
\cL^{}_{a}(x)
\=\cL^{}_{a}(\phi(x),\partial\phi(x),x)\~,
\eeq
The variational formula \e{svar} becomes
\beq
\delta S^{}_{\cV} \= \sum_{a}\int_{\cV^{a}}\dx \left[ f^{}_{a}(x) \eps(x) 
+ \jmath^{\mu}_{a}(x) d_{\mu} \eps(x) \right]\~, \label{svara}
\eeq
the global variation formula \e{globalds} becomes
\beq
\delta S^{}_{\cV} \= \eps_{0} F^{}_{\cV}\~,\qquad
{}F^{}_{\cV}\~:=~\sum_{a}F^{}_{a}\~,\qquad  
{}F^{}_{a}[\phi]\~:=\~\int_{\cV^{}_{a}}\dx\~f^{}_{a}(x)\~,
\label{globaldsa}
\eeq 
the bare Noether current \e{nnoetherhat} becomes
\beq
\jmath^{\mu}_{a}(x)\~:=\~
\frac{\partial\cL^{}_{a}(x)}{\partial\phi^{\alpha}_{\mu}(x)}Y_{0a}^{\alpha}(x)
+ \cL^{}_{a}(x) X^{\mu}_{a}(x)\~,\label{nnoethera}
\eeq
and the function \e{eff} becomes
\beq
f^{}_{a}(x)\~:=\~
\frac{\delta\cL^{}_{a}(x)}{\delta\phi^{\alpha}(x)}Y_{0a}^{\alpha}(x)
+d_{\mu}\jmath^{\mu}_{a}(x)\~.\label{effa}
\eeq
The only difference is that all quantities now carry a chart-subindex ``$a$''. 
The definition \e{dszeroa} of a global off-shell quasi-symmetry becomes
\beq
\forall \phi:\~\~\~\~
\delta S^{}_{\cV}\~\equiv\~\eps_{0} \sum_{a}G^{}_{a}\~, \qquad
G^{}_{a}[\phi]\~:=\~\int_{\cV^{}_{a}}\dx\~g^{}_{a}(x)\~, 
\label{dszeroa}
\eeq
where the integrand $g^{}_{a}$ is locally a divergence, so that the integral
$\sum_{a}G^{}_{a}$ only receives contributions from external boundaries,
\ie contributions from internal boundaries cancel pairwise. 
Then \eq{fzero} is replaced by 
\beq
\forall \phi:\~\~\~\~
\sum_{a}F^{}_{a}\~\equiv\~F^{}_{\cV}\~\equiv\~\sum_{a}G^{}_{a}\~.\label{fzeroa}
\eeq
Now define the zero functional 
\beq
Z^{}_{\cV}[\phi]\~:=\~\sum_{a}(F^{}_{a}[\phi]-G^{}_{a}[\phi])
\=\sum_{a}\int_{\cV^{}_{a}}\dx\~(f^{}_{a}-g^{}_{a})(x)\=0\~.
\label{zzeroa}
\eeq
By performing an arbitrary variation
$\delta \phi$ with support inside a
single chart $\cV^{}_{a}$ away from any boundaries, one concludes that
the Euler-Lagrange derivative vanishes identically in the interior
$\cV_{a}^{\circ}$ of $\cV_{a}^{}$,
\beq
\forall\phi \forall x\in \cV_{a}^{\circ}:\~\~\~\~ 
\frac{\delta f^{}_{a}(x)}{\delta\phi^{\alpha}(x)}\=
\frac{\delta (f^{}_{a}-g^{}_{a})(x)}{\delta\phi^{\alpha}(x)}\=0\~.
\eeq
Hence one can proceed within a single coordinate patch $\cV^{}_{a}$, as
already done in previous Sections, and prove the sought--for off--shell
Noether identity \e{offshellnihat} at the interior point $x\in\cV_{a}^{\circ}$.
All the constructions are geometrically covariant; they do not depend on the
choice of coordinate patches $\cV^{}_{a}$, or the positions of patch 
boundaries, so the Noether identity \e{offshellnihat} holds for all points
$x\in\cV$.

\section{Example: Particle with External Force}
\label{secex}

\noi
Consider the action for a non-relativistic point particle of mass $m$ moving
in one dimension, 
\beq
S^{}_{\cV}[q]\~:=\~\int_{t_{i}}^{t_{f}} \!{\rm d}t\~L(t)\~,\qquad
L(t)\~:=\~\Hf m \left(\dot{q}(t)\right)^{2}+ q(t)F(t)\~,\qquad
x^{0}\equiv t\~. \label{smodel}
\eeq
Assume that the particle experiences a given background external force $F(t)$ 
that is independent of $q$ and happens to satisfy that the total momentum 
transfer $\Delta P$ for the whole time period $[t_{i},t_{f}]$ is zero
\beq
\Delta P\=\int_{t_{i}}^{t_{f}} \!{\rm d}t\~F(t)\=0\~. 
\label{zeromomentumtransfer}
\eeq
The fixed region is in this case $\cV\!=\![t_{i},t_{f}]$. One can
write
\beq
S^{}_{\cV}[q]\=\int_{\R} \!{\rm d}t\~\hat{L}(t)\~,\qquad
\hat{L}(t)\~:=\~1^{}_{\cV}(t)L(t)\~,
\eeq
The Euler-Lagrange derivative is
\bea
\frac{\delta\hat{L}(t)}{\delta q(t)}
&=&1^{}_{\cV}(t)\frac{\delta L(t)}{\delta q(t)}
-\frac{\partial L(t)}{\partial \dot{q}(t)}\partial^{}_{0}1^{}_{\cV}(t) \cr
&=&1^{}_{\cV}(t)\left[F(t)-m\ddot{q}(t)\right]
+m\dot{q}(t)\left[ \delta(t\!-\!t_{f})-\delta(t\!-\!t_{i}) \right]\~.
\label{eomex}
\eea
The principle of stationary/least action in classical mechanics says that 
$\delta\hat{L}(t)/\delta q(t)\approx 0$ is the equations of motion for the 
system. This yields Newton's second law in the bulk,
\beq
\forall t\in \cV^{\circ}:\~\~
\frac{\delta L(t)}{\delta q(t)}\= F(t)-m\ddot{q}(t)\~\approx\~0~.
\eeq
and Neumann conditions at the boundary,
\beq
\dot{q}(t_{i})\~\approx\~0\~,\qquad \dot{q}(t_{f})\~\approx\~0\~. \label{nbc}
\eeq
Note that we here take painstaking care of representing the model 
\e{smodel} as it was mathematically given to us. The delta functions
at the boundary in \eq{eomex} may or may not reflect the physical reality. 
{}For instance, if the variational problem has additional conditions, say, a 
Dirichlet boundary condition $q(t_{i})\!=\!q_{i}$ at $t\!=\!t_{i}$, then any 
variation of $q$ must obey $\delta q(t_{i})\!=\!0$, and one will be unable to 
deduce the corresponding equation of motion for $t\!=\!t_{i}$, and therefore 
one cannot conclude the Neumann boundary condition \e{nbc} at $t\!=\!t_{i}$.
{}If the system is unconstrained at $t\!=\!t_{i}$, it will probably make more
physical sense to {\em impose} Neumann boundary condition \e{nbc} at 
$t\!=\!t_{i}$ from the very beginning, rather than to derive it as an equation
of motion. Similarly for the other boundary $t\!=\!t_{f}$.

\noi
Consider now a global variation
\beq
\delta t\=0~, \qquad
\delta q(t)\=\delta^{}_{0}q(t) \= \eps_{0}~,\label{varex}
\eeq
where $\eps_{0}$ is a global, $t$-independent infinitesimal $1$-parameter,
\ie the horizontal and vertical generators are $X^{0}(t)\!=\!0$ and 
$Y(t)\!=\!Y^{}_{0}(t)\!=\!1$, respectively. This vertical variation 
$\delta\!=\!\delta^{}_{0}$ is {\em not} necessarily a symmetry of the 
Lagrangian
\beq
\delta L(t)\=\eps_{0}F(t)\~,
\eeq
but it is a symmetry of the action
\beq
\delta S^{}_{\cV} \= \eps_{0}\Delta P\=0 \~, 
\eeq
due to the condition \e{zeromomentumtransfer}. We stress that the global 
variation \e{varex} is {\em not} necessarily a symmetry of the action for other
regions $\cU$. The bare Noether current is the momentum of the particle
\beq   
\jmath^{0}(t)
\=\frac{\partial L(t)}{\partial \dot{q}(t)}Y^{}_{0}(t)\=m\dot{q}(t)\~.
\eeq
The function
\beq
f(t)\~:=\~\frac{\delta L(t)}{\delta q(t)}Y^{}_{0}(t)+d^{}_{0}\jmath^{0}(t)
\=F(t)\~.
\eeq
from \eq{eff} can be written as a total time derivative
\beq
f(t)\=d^{}_{0}f^{0}(t)\~,
\eeq
if one defines the accumulated momentum transfer
\beq
f^{0}(t)\~:=\~\int^{t}\!{\rm d}t^{\prime}\~F(t^{\prime})\~.
\eeq
The improved Noether current is then
\beq
J^{0}(t)\~:=\~\jmath^{0}(t)-f^{0}(t)
\= m\dot{q}(t)-f^{0}(t)\~.
\eeq
The off-shell Noether identity reads
\beq
d^{}_{0}J^{0}(t)\= m\ddot{q}(t)-F(t)\=-\frac{\delta L(t)}{\delta q(t)}
Y^{}_{0}(t)\~.
\eeq

\section{Example: Particle with Fluctuating Zero-Point Energy}
\label{secex2}

\noi
Consider the action for a non-relativistic point particle of mass $m$ moving
in one dimension, 
\beq
S^{}_{\cV}[q]\~:=\~\int_{t_{i}}^{t_{f}} \!{\rm d}t\~L(t)\~,\qquad
L(t)\~:=\~T(t)-V(t)\~,\qquad
T(t)\~:=\~\Hf m \left(\dot{q}(t)\right)^{2}\~. \label{smodel2}
\eeq
Assume that the background fluctuating zero-point energy $V(t)$ is independent 
of $q$ and happens to satisfy that 
\beq
V(t_{i})\=V(t_{f})~.
\label{zeroenergychange}
\eeq
The fixed region is in this case $\cV\!\equiv\![t_{i},t_{f}]$.
The Euler-Lagrange derivative is
\beq
0 \~\approx\~ \frac{\delta L(t)}{\delta q(t)}\=-m\ddot{q}(t)\~.
\eeq
Consider now a global variation
\beq
\delta t \= -\eps_{0}~,\qquad
\delta q(t) \= 0~,\qquad
\delta^{}_{0} q(t) \= \eps_{0}\dot{q}(t)\~, \label{varex2}
\eeq
where $\eps_{0}$ is a global, $t$-independent infinitesimal $1$-parameter,
\ie the generators are $X^{0}(t)\!=\!-1$, $Y(t)\!=\!0$ and 
$Y^{}_{0}(t)\!=\!\dot{q}(t)$. This variation \e{varex2} is {\em not} 
necessarily a symmetry of the Lagrangian
\beq
\delta L(t)\=\eps_{0}\dot{V}(t)\~, \label{lquasisymex2}
\eeq
but it is a symmetry of the action
\beq
\delta S^{}_{\cV} \=\int_{t_{i}}^{t_{f}}\! {\rm d}t \left( \delta L(t)
+ L(t) d^{}_{0} \delta t \right)
\=\eps_{0}\int_{t_{i}}^{t_{f}}\! {\rm d}t\~\dot{V}(t) 
\=\eps_{0}\left[V(t_{f})\!-\!V(t_{i})\right]\=0 \~, \label{exactsym}
\eeq
due to the condition \e{zeroenergychange}. We stress that the variation
\e{varex2} is {\em not} necessarily a symmetry of the action for other 
regions $\cU$. The bare Noether current is the total energy of the particle
\beq   
\jmath^{0}(t)\~:=\~\frac{\partial L(t)}{\partial \dot{q}(t)}Y^{}_{0}(t)
+L(t)X^{0}(t)\=T(t)+V(t)\~. \label{barenoetherex2}
\eeq
The function $f(t)$ from \eq{eff} is a total time derivative of the zero-point
energy
\beq
f(t)\~:=\~\frac{\delta L(t)}{\delta q(t)}Y^{}_{0}(t)+d^{}_{0}\jmath^{0}(t)
\=\dot{V}(t)\=d^{}_{0}f^{0}(t) \label{effex2}
\eeq
if one defines $f^{0}(t)\!=\!V(t)$.
The improved Noether current is the kinetic energy
\beq
J^{0}(t)\~:=\~\jmath^{0}(t)-f^{0}(t)\= T(t)\~. \label{improvednoetherex2}
\eeq
The off-shell Noether identity reads
\beq
d^{}_{0}J^{0}(t)\=\dot{T}(t)\= m\dot{q}(t)\ddot{q}(t)
\=-\frac{\delta L(t)}{\delta q(t)}
Y^{}_{0}(t)\~.\label{noetheridex2}
\eeq
Notice that one may need to improve the bare Noether current
$\jmath^{0}(t)\to J^{0}(t)$ even in cases of an exact symmetry \e{exactsym} of
the action.

\section{Example: Quasi-Symmetry vs. Symmetry}
\label{secex3}

\noi
Here we will consider a quasi-symmetry $\delta$ of a Lagrangian $L(t)$ that 
can {\em not} be turned into a symmetry by modifying the Lagrangian 
$L(t) \rightarrow \tilde{L}(t) := L(t) + dF(t)/dt$ with a total derivative.

\noi
Let $L(t)\!=\!L(q(t),\dot{q}(t))$ be a Lagrangian that depends on position
$q(t)$ and velocity $\dot{q}(t)$, but that does {\em not} depend explicitly 
on time $t$. Consider now a global variation
\beq
\delta t \= 0~,\qquad
\delta q(t) \= \delta^{}_{0} q(t) \= \eps_{0}\dot{q}(t)\~, \label{varex3}
\eeq
where $\eps_{0}$ is a global, $t$-independent infinitesimal $1$-parameter,
\ie the generators are $X^{0}(t)\!=\!0$ and
$Y(t)\!=\!Y^{}_{0}(t)\!=\!\dot{q}(t)$. This vertical variation 
$\delta\!=\!\delta^{}_{0}$ is a quasi-symmetry of the Lagrangian
\beq
\delta L(t)\=\eps_{0}\left(\frac{\partial L(t)}{\partial q(t)} \dot{q}(t)
+\frac{\partial L(t)}{\partial \dot{q}(t)} \ddot{q}(t)\right)
\=\eps_{0}\dot{L}(t)\~, \label{lquasisymex3}
\eeq
but it is only a symmetry of the Lagrangian $\delta L(t)\!=\!0$, if $L(t)$
does also not depend on position $q(t)$ and velocity $\dot{q}(t)$, \ie if
the Lagrangian is only a constant. Thus, in order to modify the Lagrangian 
$L(t) \to \tilde{L}(t) := L(t) + dF(t)/dt$, so that the new Lagrangian 
$\delta \tilde{L}(t)\!=\!0$ has a symmetry, the old Lagrangian $L(t)$ must be 
a total derivative to begin with.

\noi
The bare Noether current $\jmath^{0}(t)$ is  
\beq   
\jmath^{0}(t)\~:=\~\frac{\partial L(t)}{\partial \dot{q}(t)}Y^{}_{0}(t)
+L(t)X^{0}(t)\=p(t)\dot{q}(t)\~. \label{barenoetherex3}
\eeq
The function $f(t)$ from \eq{eff} is a total time derivative of the Lagrangian
\beq
f(t)\~:=\~\frac{\delta L(t)}{\delta q(t)}Y^{}_{0}(t)+d^{}_{0}\jmath^{0}(t)
\=\dot{L}(t)\=d^{}_{0}f^{0}(t) \label{effex3}
\eeq
if one defines $f^{0}(t)\!=\!L(t)$.
The improved Noether current is the energy
\beq
J^{0}(t)\~:=\~\jmath^{0}(t)-f^{0}(t)\=p(t)\dot{q}(t)-L(t)\=h(t)\~. 
\label{improvednoetherex3}
\eeq
The off-shell Noether identity reads
\beq
d^{}_{0}J^{0}(t)\=\dot{h}(t)
\=-\frac{\delta L(t)}{\delta q(t)}
Y^{}_{0}(t)\~,\label{noetheridex3}
\eeq
reflecting the well-known fact that the energy $h(t)$ is conserved when the
Lagrangian does not depend explicitly on time $t$.

\vspace{0.8cm}

\noi
{\sc Acknowledgement:}~I would like to thank Bogdan Morariu for fruitful 
discussions at the Rockefeller University. The work of K.B.\ is supported by 
the Ministry of Education of the Czech Republic under the project MSM 
0021622409.

\appendix

\section{Identically Vanishing Euler-Lagrange Derivative}
\label{secapp}

\noi
We will prove in this Appendix~\ref{secapp} that an integral is a boundary 
integral if its Euler-Lagrange derivative vanishes. Consider a function 
\beq
\cL\~\in\~ {\cal F}(M\!\times\! M^{d}\!\times\! M^{d(d+1)/2}\!\times\! \cV)\~,
\qquad
\cL(x)\=\cL(\phi(x),\partial\phi(x),\partial^{2}\phi(x),x)\~,
\eeq
on the $2$-jet space. The function $\cL$ is assumed to be smooth in both
vertical and horizontal directions.

\begin{lemma}
\beq
\begin{array}{c}
{\rm Identically~vanishing~Euler~Lagrange~derivatives~of~}
\cL(x)\!=\!\cL(\phi(x),\partial\phi(x),\partial^{2}\phi(x),x): \cr
\forall \phi\forall x\in \cV:\~\~\~\~ 
\frac{\delta \cL(x)}{\delta\phi^{\alpha}(x)}\~\equiv\~0\~.
\cr\cr\Downarrow\cr\cr
\cL(x) {\rm~is~locally~a~divergence}: \cr 
\forall x^{}_{0}\in \cV \exists{\rm~local~}x^{}_{0}{\rm~neighborhood~}
\cW\subseteq \cV, \cr
\exists \Lambda^{\mu}(x)
=\Lambda^{\mu}(\phi(x),\partial\phi(x),\partial^{2}\phi(x),x)
\forall x\in \cW:\~\~ \cL(x)\=d_{\mu}\Lambda^{\mu}(x)\~.
\end{array}
\eeq
\label{lemma1}
\end{lemma}

\noi
{\sc Proof of Lemma~\ref{lemma1}}:~~
Define a region with one more dimension
\beq
\tilde{\cV}\~:=\~\cV \times [0,1]~, 
\eeq
which locally has coordinates $\tx:=(x,\lambda)$. Define the field
$\tilde{\phi}:\tilde{\cV}\!\to\! M$ as  
\beq
\tphi(\tx)\~:=\~\lambda\phi(x)\~.
\eeq
This makes sense, because the target space $M$ is star-shaped around $0$, \cf
\eq{starshaped}. Define
\beq
\tcL(\tx)\~:=\~\cL(\tphi(\tx),\partial\tphi(\tx),\partial^{2}\tphi(\tx),x)
\=\left.\cL(x)\right|_{\phi(x)\to\tphi(\tx)}\~.
\eeq
Note that $\tcL$ does not depend on $\lambda$-derivatives of the 
$\tphi$-fields, nor explicitly on $\lambda$. Thus the total derivative \wrt
$\lambda$ reads
\bea
\frac{d\tcL(\tx)}{d\lambda}
&=&\frac{\partial\tcL(\tx)}{\partial\tphi^{\alpha}(\tx)}
\frac{\partial\tphi^{\alpha}(\tx)}{\partial\lambda}
+\frac{\partial\tcL(\tx)}{\partial\tphi^{\alpha}_{\mu}(\tx)}
\frac{\partial\tphi^{\alpha}_{\mu}(\tx)}{\partial\lambda}
+\sum_{\nu\leq\mu}
\frac{\partial\tcL(\tx)}{\partial\tphi^{\alpha}_{\mu\nu}(\tx)}
\frac{\partial\tphi^{\alpha}_{\mu\nu}(\tx)}{\partial\lambda} \cr
&\equi{\e{appeom}+\e{appbiglambda}}&
\frac{\delta\tcL(\tx)}{\delta\tphi^{\alpha}(\tx)}
\frac{\partial\tphi^{\alpha}(\tx)}{\partial\lambda}
+d_{\mu} \tilde{\Lambda}^{\mu}(\tx)
\~\equi{\e{appeom}}\~d_{\mu} \tilde{\Lambda}^{\mu}(\tx)\~,\label{applambdader}
\eea
where the Euler-Lagrange derivatives vanish by assumption
\beq
\frac{\delta\tcL(\tx)}{\delta\tphi^{\alpha}(\tx)}
\~:=\~\frac{\partial\tcL(\tx)}{\partial\tphi^{\alpha}(\tx)}
-d_{\mu}\frac{\partial\tcL(\tx)}{\partial\tphi^{\alpha}_{\mu}(\tx)}
+\sum_{\nu\leq\mu}d_{\mu}d_{\nu}
\frac{\partial\tcL(\tx)}{\partial\tphi^{\alpha}_{\mu\nu}(\tx)}
\=\left.\frac{\delta\cL(x)}{\delta\phi^{\alpha}(x)}
\right|_{\phi(x)\to\tphi(\tx)}\=0\~,\label{appeom}
\eeq
and we have defined some functions
\beq
\tilde{\Lambda}^{\mu}(\tx)
\~:=\~\left(\frac{\partial\tcL(\tx)}{\partial\tphi^{\alpha}_{\mu}(\tx)}
-2\sum_{\nu\leq\mu}
d_{\nu}\frac{\partial\tcL(\tx)}{\partial\tphi^{\alpha}_{\mu\nu}(\tx)}\right)
\frac{\partial\tphi^{\alpha}(\tx)}{\partial\lambda}
+\sum_{\nu\leq\mu}d_{\nu}\left(
\frac{\partial\tcL(\tx)}{\partial\tphi^{\alpha}_{\mu\nu}(\tx)}
\frac{\partial\tphi^{\alpha}(\tx)}{\partial\lambda} \right)\~.
\label{appbiglambda}
\eeq
Hence
\beq
\cL(x)-\left.\cL(x)\right|_{\phi=0}
\=\left.\tcL(\tx)\right|_{\lambda=1}-\left.\tcL(\tx)\right|_{\lambda=0}
\=\int_{0}^{1}\! {\rm d}\lambda\frac{d\tcL(\tx)}{d\lambda}
\\~\equi{\e{applambdader}}\~
d_{\mu}\int_{0}^{1}\! {\rm d}\lambda\~\tilde{\Lambda}^{\mu}(\tx)\~.
\label{ldiff}
\eeq
On the other hand, the lower boundary 
\beq
h(x)\~:=\~\left.\cL(x)\right|_{\phi=0}
\eeq
in \eq{ldiff} does not depend on $\phi$, so one can, \eg locally pick a 
coordinate $t\!\equiv\!x^{0}$, so that $x^{\mu}\!=\!(t,\vec{x})$, and define
\beq
H^{0}(x):=\int^{t} \! {\rm d}t^{\prime}~h(t^{\prime},\vec{x})\~, \qquad
0\=H^{1}\=H^{2}\=\ldots\=H^{d-1}\~.
\eeq
Then $h(x)\!=\!\partial_{\mu}H^{\mu}(x)$ is locally a divergence.
Altogether, this implies that $\cL(x)$ is locally a divergence.
\proofbox

\noi
{\sc Remark}:~~It is easy to check that the opposite arrow ``$\Uparrow$'' in 
Lemma~\ref{lemma1} is also true. The Lemma~\ref{lemma1} can be generalized to 
$n$-jets, for any $n=1,2,3,\ldots$, using essentially the same proof technique.
We have focused on the $n\!=\!2$ case, since this is the case that is needed 
in the proof of Noether's first Theorem, \cf \eq{fdivhat}. The fact that the 
$n\!=\!2$ case is actually needed for the physically relevant case, where the
Lagrangian density depends on up to first order derivatives of the fields, is 
often glossed over in standard textbooks on classical mechanics. By (a 
dualized version of) the Poincar\'e Lemma, it follows that the local functions
$\Lambda^{\mu} \to \Lambda^{\mu} + d^{}_{\nu}\Lambda^{\nu\mu}$ are unique up to 
antisymmetric improvement terms $\Lambda^{\nu\mu}\!=\! -\Lambda^{\mu\nu}$,
see \eg \Ref{bbh00}.

\section{Gauging a Global $u(1)$ Quasi-Symmetry}
\label{secgauge}

\noi
A global quasi-symmetry $\delta$ from \eq{variation} is by definition promoted
to a {\em gauge quasi-symmetry} if the variation $\delta S^{}_{\cV}$ of the
action in \eq{svar} is a boundary integral for arbitrary $x$-dependent
$\eps(x)$. Noether's second Theorem \cite{noether18} states that a gauge
quasi-symmetry $\delta$ implies an off-shell conservation law and an off-shell
Noether identity, \ie 
\beq
0\~\equiv\~d_{\mu}J^{\mu}(x)\~\equiv\~
-\frac{\delta\cL(x)}{\delta\phi^{\alpha}(x)}Y_{0}^{\alpha}(x)\~.
\label{offshellnihat2}
\eeq
As we shall see in \eq{gaugetransfgaugedlagr}, it is often possible to gauge a
global $u(1)$ quasi-symmetry $\delta$ by introducing an Abelian {\em gauge
potential} $A_{\mu}\!=\!A_{\mu}(x)$ with infinitesimal Abelian {\em gauge
transformation}
\beq
\delta A^{}_{\mu} \= \partial^{}_{\mu}\eps\~,\label{gaugetransf}
\eeq
and adding certain terms to the Lagrangian density $\cL$ that vanish for
$A\to0$. The Abelian field strength
\beq
F^{}_{\mu\nu}\~:=\~\partial^{}_{\mu}A^{}_{\nu} - (\mu\leftrightarrow\nu) 
\eeq
is gauge invariant $\delta F^{}_{\mu\nu}\!=\!0$.
In this appendix, we specialize to the case where the horizontal
generator $X^{\mu}$ vanishes, and where the vertical generator $Y^{\alpha}_{0}$ 
does not depend on derivatives $\partial\phi$,
\beq
X^{\mu}(x) \= 0\~, \qquad Y^{\alpha}_{0}(x) \= Y^{\alpha}_{0}(\phi(x),x)\~.
\label{minassumption}
\eeq
Assumption \e{minassumption} is made in order for the sought-for gauged
Lagrangian density $\cL^{\gauged}$ to be minimally coupled, \cf
\eq{gaugedlagr}. It is useful to first introduce a bit of notation. The {\em
jet-prolongated vector field} $\hY^{}_{0}$ is defined as
\beq
\hY^{}_{0}\~:=\~J^{\bullet}Y^{}_{0} 
\=Y^{\alpha}_{0}\frac{\partial}{\partial\phi^{\alpha}}
+ d^{}_{\mu}Y^{\alpha}_{0}\~\frac{\partial}{\partial\phi^{\alpha}_{\mu}} 
+\sum_{\mu\leq\nu} d^{}_{\mu} d^{}_{\nu}Y^{\alpha}_{0}\~
\frac{\partial}{\partial\phi^{\alpha}_{\mu\nu}} + \ldots\~.\label{jetprol}
\eeq
The  jet-prolongated vector field $\hY^{}_{0}$ and the total derivative 
$d^{}_{\mu}$ commute $[d^{}_{\mu},\hY^{}_{0}]\!=\!0$. The {\em covariant
derivative} $D^{}_{\mu}$ is defined as
\beq
D^{}_{\mu}\~:=\~d^{}_{\mu}
- A^{}_{\mu}Y^{\alpha}_{0}\frac{\partial}{\partial\phi^{\alpha}}\~. 
\label{covderiv}
\eeq
The characteristic feature of the covariant derivative
$D^{}_{\mu}\phi^{\alpha}=d^{}_{\mu}\phi^{\alpha}- A^{}_{\mu}Y^{\alpha}_{0}$
is that it behaves covariantly under the gauge transformation $\delta$,
\beq
\delta D^{}_{\mu}\phi^{\alpha} 
\= d^{}_{\mu}\delta\phi^{\alpha} - Y^{\alpha}_{0}\delta A^{}_{\mu}
- A^{}_{\mu}\delta Y^{\alpha}_{0}
\= d^{}_{\mu}(\eps Y^{\alpha}_{0}) - Y^{\alpha}_{0}\partial^{}_{\mu}\eps
- A^{}_{\mu}\frac{\partial Y^{\alpha}_{0}}{\partial\phi^{\beta}}Y^{\beta}_{0}\eps
\= \eps D^{}_{\mu}Y^{\alpha}_{0}\~. \label{gaugetransfcovderiv}
\eeq 
The {\em minimal} extension $\tilde{h}(x)$ (which in this
Appendix~\ref{secgauge} is notationally denoted with a tilde ``$\sim$'') of a
function
\beq
h(x)\=h(\phi(x),\partial\phi(x),A(x),F(x),x)\~, \label{age}  
\eeq
is defined by replacing partial derivatives $\partial^{}_{\mu}$ with covariant
derivatives $D^{}_{\mu}$, \ie
\beq
\tilde{h}(x)\~:=\~h(\phi(x),D\phi(x),A(x),F(x),x)\~.\label{agetilde} 
\eeq
Here it is important that the $h$ function in \eq{age} does {\em not} depend on
higher $x$-derivatives of $\phi$. (A minimal extension $\tilde{h}$ of a
function $h$, that depend on higher $x$-derivatives of $\phi$, is only
well-defined if the field strength $F^{}_{\mu\nu}$ vanishes, so that the
covariant derivatives $D^{}_{\mu}$ commute.) Assumption \e{minassumption}
implies that
\bea
\left(d^{}_{\mu}\hY^{}_{0}\phi^{\alpha}\right)^{\sim} 
&=&\left(d^{}_{\mu}Y^{\alpha}_{0}\right)^{\sim} 
\=\left(\partial^{}_{\mu}Y^{\alpha}_{0}
+\frac{\partial Y^{\alpha}_{0}}{\partial\phi^{\beta}}
\partial^{}_{\mu}\phi^{\beta} \right)^{\sim}
\= \partial^{}_{\mu}Y^{\alpha}_{0}
+\frac{\partial Y^{\alpha}_{0}}{\partial\phi^{\beta}}D^{}_{\mu}\phi^{\beta}  \cr 
&=&D^{}_{\mu}Y^{\alpha}_{0}
\=d^{}_{\mu}Y^{\alpha}_{0}
-A^{}_{\mu} Y^{\beta}_{0}\frac{\partial Y^{\alpha}_{0}}{\partial\phi^{\beta}}
\=d^{}_{\mu}\hY^{}_{0}\phi^{\alpha}
-A^{}_{\mu}\hY^{}_{0} Y^{\alpha}_{0} \cr
&=&\hY^{}_{0}D^{}_{\mu}\phi^{\alpha}\~.
\label{helpinghand}
\eea
More generally, assumption \e{minassumption} implies that the jet-prolongated
vector field $\hY^{}_{0}$ and the minimal extension ``$\sim$'' commute in the
sense that if $h$ is a function of type \e{age}, then $\hY^{}_{0}h$ is also a
function of type \e{age}, and its minimal extension is
\beq
\left(\hY^{}_{0}h\right)^{\sim} 
\=\left(\frac{\partial h}{\partial\phi^{\alpha}}Y^{\alpha}_{0}
+ \frac{\partial h}{\partial\phi^{\alpha}_{\mu}}d^{}_{\mu}Y^{\alpha}_{0}
\right)^{\sim} 
\~\equi{\e{helpinghand}}\~
\frac{\partial \tilde{h}}{\partial\phi^{\alpha}}Y^{\alpha}_{0}
+ \frac{\partial\tilde{h}}{\partial D^{}_{\mu}\phi^{\alpha}}
D^{}_{\mu}Y^{\alpha}_{0} 
\~\equi{\e{helpinghand}}\~\hY^{}_{0}\tilde{h}\~.
\eeq
{}Furthermore, the gauge transformation $\delta\tilde{h}$ of the minimal
extension $\tilde{h}$ can be calculated \wthot jet-prolongated vector field
$\hY^{}_{0}$ as
\beq
\delta \tilde{h} 
\= \frac{\partial \tilde{h}}{\partial\phi^{\alpha}}\delta\phi^{\alpha}
+ \left(\frac{\partial h}{\partial\phi^{\alpha}_{\mu}} \right)^{\sim} 
\delta D^{}_{\mu}\phi^{\alpha}
+ \left(\frac{\partial h}{\partial A^{}_{\mu}} \right)^{\sim}\delta A^{}_{\mu}
\= \left( \eps\hY^{}_{0}h
+\frac{\partial h}{\partial A^{}_{\mu}}\partial^{}_{\mu}\eps \right)^{\sim} \~.
\label{gaugedth}
\eeq
In particular, it follows from assumption \e{minassumption} that the function 
$f\!=\!\hY^{}_{0}\cL$ from \eq{eff} is a function of type \e{age},
\ie $f$ can not depend on higher $x$-derivatives of the field $\phi$,
 \beq
f(x)\=f(\phi(x),\partial\phi(x),x)\~.\label{eff2}
\eeq
Equation \e{eff2} and Appendix~\ref{secapp} imply, in turn, that the local
function $f^{\mu}(x)\!=\!f^{\mu}(\phi(x),\partial\phi(x),x)$ from \eq{fdivhat}
must also be of type \e{age}, and have derivatives
\beq
\frac{\partial f^{\mu}}{\partial\phi^{\alpha}_{\nu}}
\= -(\mu\leftrightarrow\nu) \label{effmunuantisym}
\eeq
that are $\mu\leftrightarrow\nu$ antisymmetric. The local function
$f^{\mu} \to f^{\mu} + d^{}_{\nu}f^{\nu\mu}$ is unique up to antisymmetric
improvement terms $f^{\nu\mu}\!=\! -f^{\mu\nu}$. We will furthermore assume
that 
\beq
f^{\mu}~{\rm is~globally~defined}, \label{effmuglobal}
\eeq
and that $f^{\mu}$ has been chosen so that 
\beq
\frac{\partial f^{\mu}}{\partial\phi^{\alpha}_{\nu}}Y^{\alpha}_{0}
\= (\mu\leftrightarrow\nu)\~. \label{effmunusym}
\eeq
The latter assumption \e{effmunusym} together with \eq{effmunuantisym} imply
that
\beq
\frac{\partial f^{\mu}}{\partial\phi^{\alpha}_{\nu}}Y^{\alpha}_{0}
\=0\~. \label{effmununull}
\eeq
The (minimally coupled) {\em gauged Lagrangian density} $\cL^{\gauged}$ is now
defined as
\beq
\cL^{\gauged}\~:=\~\left(\cL + A^{}_{\mu}f^{\mu} \right)^{\sim}
\= \tcL + A^{}_{\mu}\tilde{f}^{\mu} \~. \label{gaugedlagr} 
\eeq
The gauge transformation $\delta\cL^{\gauged}$ of $\cL^{\gauged}$ can be written
as a divergence
\bea
\delta \cL^{\gauged}&\equi{\e{gaugedth}}&
\left(\eps \hY^{}_{0}\cL + f^{\mu}\partial^{}_{\mu}\eps  
+  A^{}_{\mu}\eps\hY^{}_{0}f^{\mu} \right)^{\sim} \cr
&=& \left( d^{}_{\mu}(\eps f^{\mu})  
+\eps A^{}_{\mu}(\frac{\partial f^{\mu}}{\partial\phi^{\alpha}}Y^{\alpha}_{0}
+ \frac{\partial f^{\mu}}{\partial\phi^{\alpha}_{\nu}}d^{}_{\nu}Y^{\alpha}_{0}
 )\right)^{\sim} \cr
&=& \partial^{}_{\mu}(\eps \tilde{f}^{\mu})
+\eps\frac{\partial\tilde{f}^{\mu}}{\partial\phi^{\alpha}}D^{}_{\mu}\phi^{\alpha} 
+\eps A^{}_{\mu}\left(
\frac{\partial\tilde{f}^{\mu}}{\partial\phi^{\alpha}}Y^{\alpha}_{0}
+ \frac{\partial\tilde{f}^{\mu}}{\partial D^{}_{\nu}\phi^{\alpha}}
D^{}_{\nu}Y^{\alpha}_{0}\right)\cr
&\equi{\e{effmunuantisym}}&\partial^{}_{\mu}(\eps \tilde{f}^{\mu})
+\eps\frac{\partial\tilde{f}^{\mu}}{\partial\phi^{\alpha}}
\partial^{}_{\mu}\phi^{\alpha} 
-\eps A^{}_{\mu}\frac{\partial\tilde{f}^{\nu}}{\partial D^{}_{\mu}\phi^{\alpha}}
d^{}_{\nu}Y^{\alpha}_{0} \cr
&\equi{\e{dtf}}& 
d^{}_{\mu}(\eps \tilde{f}^{\mu})\~, \label{gaugetransfgaugedlagr} 
\eea
because
\bea
d^{}_{\mu}\tilde{f}^{\mu} - \partial^{}_{\mu}\tilde{f}^{\mu}
-\frac{\partial\tilde{f}^{\mu}}{\partial\phi^{\alpha}}
\partial^{}_{\mu}\phi^{\alpha}
&=&\frac{\partial\tilde{f}^{\mu}}{\partial D_{\nu}\phi^{\alpha}}
d^{}_{\mu}D^{}_{\nu}\phi^{\alpha}
\~\equi{\e{effmunuantisym}}\~
-\frac{\partial\tilde{f}^{\mu}}{\partial D_{\nu}\phi^{\alpha}}
d^{}_{\mu}\left( A^{}_{\nu}Y^{\alpha}_{0} \right) \cr
&\equi{\e{effmunuantisym}}&
-\frac{\partial\tilde{f}^{\mu}}{\partial D_{\nu}\phi^{\alpha}}
\left(\Hf F^{}_{\mu\nu}Y^{\alpha}_{0}+A^{}_{\nu}d^{}_{\mu}Y^{\alpha}_{0}\right)\cr
&\equi{\e{effmunusym}}& 
-\frac{\partial\tilde{f}^{\mu}}{\partial D_{\nu}\phi^{\alpha}}
A^{}_{\nu}d^{}_{\mu}Y^{\alpha}_{0}\~. \label{dtf}  
\eea
Equation \e{gaugetransfgaugedlagr} shows that the gauge transformation $\delta$
is a gauge quasi-symmetry of the (minimally coupled) gauged action 
\beq
S^{\gauged}_{\cV}[\phi,A]\~:=\~\int_{\cV}\dx\~\cL^{\gauged}(x)\~,
\label{gaugedaction}
\eeq
which was the goal of Appendix~\ref{secgauge}.

\begin{theorem}
If an $S^{}_{\cV}$ action \e{action0} has a global quasi-symmetry 
(\ref{variation}, \ref{globalvar}, \ref{dszero}) of the form 
(\ref{minassumption}, \ref{effmuglobal}, \ref{effmunusym}),
then the (minimally coupled) gauged action \e{gaugedaction}
has a corresponding gauge quasi-symmetry. 
\end{theorem}

\noi
{}Finally, let us try to justify assumption \e{effmunusym}, which was only used
in the last equality of \eq{dtf}. Notice that the function
$f\!=\!\hY^{}_{0}\cL$ depends linearly on $\cL$, so we may argue term by term
in $\cL$. {}Firstly, in the special case where the Lagrangian density
$\cL\!=\!d^{}_{\mu}\Lambda^{\mu}$ is locally a divergence, the equations of
motion $\delta\cL/\delta\phi^{\alpha}$ vanish identically, \cf
Appendix~\ref{secapp}, and we may pick $f^{\mu}\!=\!\jmath^{\mu}$ globally as
the bare Noether current
$\jmath^{\mu}\!:=\!Y^{\alpha}_{0}\partial\cL / \partial\phi^{\alpha}_{\mu}$, 
which clearly satisfies condition \e{effmunusym}. Secondly, in the general case
with general local $f^{\mu}$, and under the additional assumption of a homotopy
inverse to the jet prolongation $\hY^{}_{0}$, there exists a local
$\Lambda^{\mu}$ such that $f^{\mu}\!=\!\hY^{}_{0}\Lambda^{\mu}$.
Since $[d^{}_{\mu},\hY^{}_{0}]\!=\!0$, we have 
$\hY^{}_{0}d^{}_{\mu}\Lambda^{\mu}\!=\!d^{}_{\mu}\hY^{}_{0}\Lambda^{\mu}
\!=\!d^{}_{\mu}f^{\mu}\!=\!f$. Because we have already discussed the special
case of a local divergence, we may subtract the local divergence 
$d^{}_{\mu}\Lambda^{\mu}$ from $\cL$, so that the remaining Lagrangian density 
$\cL^{\prime}=\cL-d^{}_{\mu}\Lambda^{\mu}$ has
vanishing $f$-function $f^{\prime}\!=\!d^{}_{\mu}f^{\prime\mu}$, because 
$f^{\prime}=\hY^{}_{0}\cL^{\prime}
=\hY^{}_{0}\cL-\hY^{}_{0}d^{}_{\mu}\Lambda^{\mu}=f-f=0$.
Thus we may pick the remaining $f^{\mu}$-function globally as 
$f^{\prime\mu}\!=\!0$, which clearly also satisfies condition \e{effmunusym}.

\noi
If we consider a point $x_{0}$, where the vertical vector field
$Y^{}_{0}(x_{0})\!\neq\!0$ does not vanish, it is possible to locally 
stratify $Y^{}_{0}$, \ie by changing target space coordinates
$\phi^{\alpha}$, so that vertical vector field
$Y^{}_{0}\!=\!\partial/\partial \phi^{1}$ (and hence the whole jet prolongation
$\hY^{}_{0}\!=\!Y^{}_{0}\!=\!\partial/\partial \phi^{1}$) is just a
differentiation \wrt a single coordinate $\phi^{1}$, the homotopy inverse
exists and is just an integration \wrt $\phi^{1}$.

\noi
In fact, these arguments show under the assumption \e{minassumption}, that
locally (away from singular points  $x_{0}$ with $Y^{}_{0}(x_{0})\!=\!0$),  
it is possible to enhance a global quasi-symmetry into a genuine global 
symmetry with vanishing function $f\!\equiv\!0$ by adding a local divergence
term $d^{}_{\mu}\Lambda^{\mu}$ to the Lagrangian density 
$\cL \to \cL + d^{}_{\mu}\Lambda^{\mu}$.

\end{document}